\documentclass[twocolumn,aps,showpacs]{revtex4}
\usepackage{amssymb}
\usepackage{epsf}

\begin{document}

\title{Landau level degeneracy and quantum Hall effect in a graphite bilayer}
\author{Edward McCann and Vladimir I. Fal'ko}
\affiliation{Department of Physics, Lancaster University, Lancaster, LA1 4YB, United
Kingdom}

\begin{abstract}
We derive an effective two-dimensional Hamiltonian to describe the
low energy electronic excitations of a graphite bilayer, which
correspond to chiral quasiparticles with a parabolic dispersion
exhibiting Berry phase $2\pi$. Its high-magnetic-field Landau
level spectrum consists of almost equidistant groups of four-fold
degenerate states at finite energy and eight zero-energy states.
This can be translated into the Hall conductivity dependence on
carrier density, $\sigma _{xy}(N)$, which exhibits plateaus at
integer values of $4e^{2}/h$ and has a \textquotedblleft
double\textquotedblright\ $8e^{2}/h$ step between the hole and
electron gases across zero density, in contrast to $(4n+2)e^{2}/h$
sequencing in a monolayer.
\end{abstract}

\maketitle

For many decades, the electronic properties of a graphite monolayer have
attracted theoretical interest due to a Dirac-type spectrum of charge
carriers \cite{mcc56,d+m84,haldane88,cham96,khv01,sta05} in this gapless
semiconductor \cite{dressel02}. Recently Novoselov \textit{et al.} \cite%
{novo04} fabricated ultra-thin graphitic devices including monolayer
structures. This was followed by further observations \cite%
{novo05pnas,novo05,zhang05} of the classical and quantum Hall effects (QHE)
in such systems confirming the expectations \cite{haldane88} of an unusual
phase of Shubnikov de Haas oscillations and QHE plateaus sequencing, as
manifestations of a peculiar magneto-spectrum of chiral Dirac-type
quasiparticles containing a Landau level at zero energy \cite{mcc56}.

In this Letter we show that quasiparticles in a graphite bilayer
display even more intriguing properties including a peculiar
Landau level (LL) spectrum: these are chiral quasiparticles
exhibiting Berry phase $2\pi$, with a dominantly parabolic
dispersion and a double-degenerate zero-energy LL incorporating
two different orbital states with the same energy. Taking into
account spin and valley degeneracies, the zero-energy LL in a
bilayer is 8-fold degenerate, as compared to the 4-fold degeneracy
of other bilayer states and the 4-fold degeneracy of all LLs in a
monolayer. The structure and degeneracies of the Landau level
spectrum in a bilayer determine a specific
sequencing of plateaus in the density dependence of the QHE conductivity $%
\sigma _{xy}(N)$ which is distinguishably different from that of Dirac-type
quasiparticles in a graphite monolayer and of non-chiral carriers in
conventional semiconductor structures.

\begin{figure}[t]
\centerline{\epsfxsize=1.0\hsize \epsffile{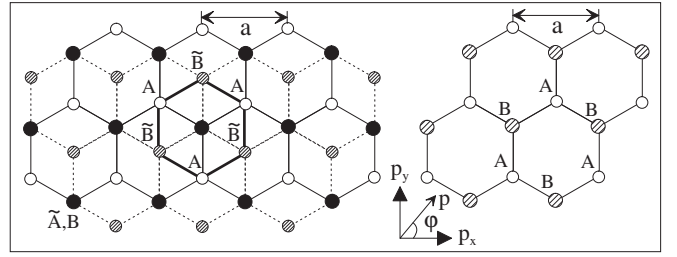}}
\caption{Left: schematic of the bilayer lattice (bonds in the bottom layer $%
A,B$ are indicated by solid lines and in the top layer $\tilde{A},\tilde{B}$
by dashed lines) containing four sites in the unit cell: $A$ (white
circles), $\tilde{B}$ (hashed), $\tilde{A}B$ dimer (solid). Right: the
lattice of a monolayer.}
\label{fig:1}
\end{figure}

We model a graphite bilayer as two coupled hexagonal lattices including
inequivalent sites $A,B$ and $\tilde{A},\tilde{B}$ in the bottom and top
layers, respectively. These are arranged according to Bernal ($\tilde{A}$-$B$%
) stacking \cite{dressel02,trick92,yosh96}, as shown in Fig.~\ref{fig:1}. A
lattice with such symmetry supports a degeneracy point at each of two
inequivalent corners, $K$ and $\tilde{K}$, of the hexagonal Brillouin zone
\cite{kpoints}, which coincide with the Fermi point in a neutral structure
and determine the centers of two valleys of a gapless spectrum. At the
degeneracy point, electron states on inequivalent ($A$/$B$ or $\tilde{A}$/$%
\tilde{B}$) sublattices in a single layer are decoupled, whereas interlayer
coupling $\gamma _{\tilde{A}B}\equiv \gamma _{1}$ forms 'dimers' from pairs
of $\tilde{A}$-$B$ orbitals in a bilayer [solid circles in Fig.~\ref{fig:1}%
], thus leading to the formation of high energy bands \cite{trick92,yosh96}.

The low energy states of electrons are described by%
\begin{eqnarray}
{\hat{H}}_{2} &=&-\frac{1}{2m}\left(
\begin{array}{cc}
0 & \left( {\pi }^{\dag }\right) ^{2} \\
{\pi ^{2}} & 0%
\end{array}%
\right) +{\hat{h}}_{w}+{\hat{h}}_{a};  \label{heff1} \\
{\hat{h}}_{w} &=&\xi v_{3}\left(
\begin{array}{cc}
0 & {\pi } \\
{\pi }^{\dag } & 0%
\end{array}%
\right) ,\;\;\;\;\mathrm{where}\;\;\;\;\;{\pi }={p}_{x}+i{p}_{y};  \nonumber
\\
{\hat{h}}_{a} &=&\xi u\left[ {\textstyle\frac{1}{2}}\left(
\begin{array}{cc}
1 & 0 \\
0 & -1%
\end{array}%
\right) -\frac{v^{2}}{\gamma _{1}^{2}}\left(
\begin{array}{cc}
{\pi }^{\dag }{\pi } & 0 \\
0 & -{\pi \pi }^{\dag }%
\end{array}%
\right) \right] .  \nonumber
\end{eqnarray}%
The effective Hamiltonian ${\hat{H}}_{2}$ operates in the space of
two-component wave functions $\Phi $ describing electronic amplitudes on $A$
and $\tilde{B}$ sites and it is applicable within the energy range $%
|\varepsilon |<\frac{1}{4}\gamma _{1}$. In the valley $K$, $\xi =+1$, we
determine $\Phi _{\xi =+1}=(\phi (A),\phi (\tilde{B}))$, whereas in the
valley $\tilde{K}$, $\xi =-1$ and the order of components is reversed, $\Phi
_{\xi =-1}=(\phi (\tilde{B}),\phi (A))$. {Here, we }take into account two
possible ways of $A\rightleftharpoons \tilde{B}$ hopping: via the dimer
state (the main part) or due to a weak direct $A\tilde{B}$ coupling, $\gamma
_{A\tilde{B}}\equiv \gamma _{3}\ll \gamma _{\tilde{A}B}$ (the term ${\hat{h}}%
_{w}$). They determine the mass $m=\gamma _{1}/2v^{2}$ and velocity $%
v_{3}=\left( \sqrt{3}/2\right) a\gamma _{A\tilde{B}}/\hbar $. Other weaker
tunneling processes \cite{dressel02} are neglected. The term ${\hat{h}}_{a}$
takes into account a possible asymmetry between top and bottom layers (thus
opening a mini-gap $\sim u$).

For comparison, the monolayer Hamiltonian \cite{d+m84},
\[
{\hat{H}}_{1}=\xi v\left(
\begin{array}{cc}
0 & {\pi }^{\dag } \\
{\pi } & 0%
\end{array}%
\right) \equiv \xi v\left( \sigma _{x}{p}_{x}+\sigma _{y}{p}_{y}\right) ,
\]%
is dominated by nearest neighbor intralayer hopping $\gamma _{AB}=\gamma
_{BA}\equiv \gamma _{0}\gg \gamma _{1}$, so that $v=\left( \sqrt{3}/2\right)
a\gamma _{AB}/\hbar $. For equivalent parameters in bulk graphite \cite%
{dressel02}, $v_{3}\ll v$ in Eq.~(\ref{heff1}). Thus, the linear
term ${\hat{h}}_{w}$, which is similar to ${\hat{H}}_{1}$, is
relevant only for very small electron momenta (i.e., in an
electron gas with a small density at a very low magnetic field)
whereas the energy spectrum within the interval
${\textstyle\frac{1}{2}}\gamma _{1}(v_{3}/v)^{2}<|\varepsilon
|<{\textstyle\frac{1}{4}}\gamma _{1}$ is dominated by the first
term \cite{H2} in ${\hat{H}}_{2}$ producing a dispersion
$|\epsilon |=p^{2}/2m$, which contrasts with $|\epsilon |=vp$ in a
monolayer. ${\hat{H}}_{1}$ and ${\hat{H}}_{2}$ form a family of
Hamiltonians ${\hat{H}}_{J}=\xi ^{J}f(|p|)\,\mathbf{\sigma }\cdot
\mathbf{n}$ describing particles which are chiral in the
sublattice space, where $\mathbf{n}=\mathbf{l}_{x}\cos (J\varphi
)+\mathbf{l}_{y}\sin (J\varphi )$ for $\mathbf{p/}p=(\cos \varphi
, \sin \varphi )$ [$\pi =pe^{i\varphi }$], and the degree of
chirality is $J=1$ in a monolayer and $J=2$ in the bilayer. It is
interesting to notice that quasiparticles described by the
Hamiltonians ${\hat{H}}_{J}$ acquire a Berry phase $J\pi$ upon an
adiabatic propagation along a closed orbit, thus charge carriers
in a bilayer are Berry phase $2\pi$  quasiparticles, in contrast
to Berry phase $\pi$ particles in the monolayer of graphene
\cite{haldane88}. According to the inverted definition of
sublattice components (for which we reserve $2\times 2$ Pauli
matrices $\sigma _{i}$ \cite{footnoteSpaceRev}) of the wave
functions $\Phi _{\xi =+1}$ and $\Phi _{\xi =-1}$, quasiparticles
in different valleys, $\xi =\pm 1$ have effectively the opposite
chirality. Also, the existence of two valleys is crucial for the
time-reversal symmetry of the chiral Hamiltonians. In application
to ${\hat{H}}_{2}$, time reversal is described by $(\Pi
_{1}\otimes \sigma _{x}){\hat{H}}^{\ast }\left(
{\mathbf{p},B,{u}}\right) (\Pi _{1}\otimes \sigma
_{x})={\hat{H}}\left( -\mathbf{p},-B,{u}\right) $, where $\Pi
_{1}$ swaps $\xi =+1$ and $\xi =-1$ in valley space
\cite{footnoteSpaceRev}.

A microscopic analysis leading to the bilayer Hamiltonian ${\hat{H}}_{2}$
uses the tight-binding model of graphite and the Slonczewski-Weiss-McClure
parameterization \cite{dressel02} of relevant couplings. We represent the
Hamiltonian near the centers of the valleys in a basis with components
corresponding to atomic sites $A,\tilde{B},\tilde{A},B$ in the valley $K$
\cite{kpoints} and to $\tilde{B},A,B,\tilde{A}$ in the valley $\tilde{K}$,
and distinguish between on-site energies, $\pm {\textstyle\frac{1}{2}}u$ in
the two layers,
\[
\mathcal{H}=\xi \left(
\begin{array}{cccc}
\frac{1}{2}{u} & v_{3}{\pi } & 0 & v{\pi }^{\dag } \\
v_{3}{\pi }^{\dag } & -\frac{1}{2}{u} & v{\pi } & 0 \\
0 & v{\pi }^{\dag } & -\frac{1}{2}{u} & \xi \gamma _{1} \\
v{\pi } & 0 & \xi \gamma _{1} & \frac{1}{2}{u}%
\end{array}%
\right) ;\!\!\!\quad \left\{
\begin{array}{l}
{\pi }={p}_{x}+i{p}_{y}, \\
\mathbf{p}=-i\hbar \mathbf{\nabla }-e\mathbf{A,} \\
\mathbf{B}={\mathrm{rot}}\mathbf{A,} \\
\lbrack \pi ,{\pi }^{\dag }]=2\hbar eB.%
\end{array}%
\right.
\]

The Hamiltonian $\mathcal{H}$ determines the following spectrum of electrons
in a bilayer at zero magnetic field. There are four valley-degenerate bands,
$\epsilon _{\alpha }^{\pm }(\mathbf{p})$, $\alpha =1,2$, with
\begin{eqnarray*}
\epsilon _{\alpha }^{2} &=&\frac{\gamma _{1}^{2}}{2}+\frac{{u}^{2}}{4}%
+\left( v^{2}+\frac{v_{3}^{2}}{2}\right) p^{2}+\left( -1\right) ^{\alpha }%
\bigg[\frac{\left( \gamma _{1}^{2}-v_{3}^{2}p^{2}\right) ^{2}}{4} \\
&&+v^{2}p^{2}\left[ \gamma _{1}^{2}+{u}^{2}+v_{3}^{2}p^{2}\right] +2\xi
\gamma _{1}v_{3}v^{2}p^{3}\cos 3\varphi \bigg]^{1/2},
\end{eqnarray*}%
where $\epsilon _{2}$ describes the higher-energy ($\tilde{A}B$ dimer) bands.

The dispersion $\epsilon _{1}(p)$ describes low energy bands. In the
intermediate energy range, ${\textstyle\frac{1}{2}}\gamma _{1}\left(
v_{3}/v\right) ^{2},u<\left\vert \epsilon _{1}\right\vert <\gamma _{1}$, it
can be approximated with
\begin{equation}
\epsilon _{1}^{\pm }\approx \pm {\textstyle\frac{1}{2}}\gamma _{1}\left[
\sqrt{1+4v^{2}p^{2}/\gamma _{1}^{2}}-1\right] .  \label{epone}
\end{equation}%
This corresponds to the effective mass for electrons near the Fermi energy
in a 2D gas with density $N$, $m_{c}=p/(\partial \epsilon _{1}/\partial
p)=\left( \gamma _{1}/2v^{2}\right) \sqrt{1+4\pi \hbar ^{2}v^{2}N/\gamma
_{1}^{2}}$. The relation in Eq.~(\ref{epone}) interpolates between a linear
spectrum $\epsilon _{1}\approx vp$ at high momenta and a quadratic spectrum $%
\epsilon _{1}\approx p^{2}/2m$, where $m=\gamma _{1}/2v^{2}$. Such a
crossover happens at $p\approx \gamma _{1}/2v$, which corresponds to the
carrier density $N^{\ast }\approx \gamma _{1}^{2}/\left( 4\pi \hbar
^{2}v^{2}\right) $. The experimental graphite values \cite{dressel02,novo05}
give $N^{\ast }\approx 4.36\times 10^{12}cm^{-2}$, whereas the dimer band $%
\epsilon _{2}$ becomes occupied only if the carrier density exceeds $%
N^{(2)}\approx 2\gamma _{1}^{2}/\left( \pi \hbar ^{2}v^{2}\right) \approx
8N^{\ast }\approx 3.49\times 10^{13}cm^{-2}$. The estimated effective mass $%
m $ is light: $m\approx 0.054m_{e}$ using the bulk graphite values \cite%
{dressel02,novo05}.

The $4\times 4$ Hamiltonian $\mathcal{H}$ contains information about the
higher energy band $\epsilon _{2}$, and, therefore, is not convenient for
the analysis of transport properties of a bilayer which are formed by
carriers in the low energy band $\epsilon _{1}$. We separate $\mathcal{H}$
into $2\times 2$ blocks, where the upper left diagonal block is $%
H_{11}\equiv \xi ({\textstyle\frac{1}{2}}{u}\sigma _{z}+v_{3}[\sigma _{x}{p}%
_{x}-\sigma _{y}{p}_{y}])$, the lower right diagonal block is $H_{22}=-{%
\textstyle\frac{1}{2}}\xi {u}\sigma _{z}+\gamma _{1}\sigma _{x}$, and the
off-diagonal blocks are $H_{21}=H_{12}=v\xi (\sigma _{x}{p}_{x}+\sigma _{y}{p%
}_{y})$. Then, we take the $4\times 4$ Green function determined by $%
\mathcal{H}$, evaluate the block $G_{11}$ related to the lower-band states,
and use it to identify the effective low-energy bilayer Hamiltonian ${\hat{H}%
}_{2}$. Using $G_{\alpha \alpha }^{\left( 0\right) }=\left( H_{\alpha \alpha
}-\varepsilon \right) ^{-1}$, we write
\[
G=\left(
\begin{array}{cc}
G_{11} & G_{12} \\
G_{21} & G_{22}%
\end{array}%
\right) =\left(
\begin{array}{cc}
G_{11}^{(0)-1} & H_{12} \\
H_{21} & G_{22}^{(0)-1}%
\end{array}%
\right) ^{-1}\!\!\!\!\equiv \left( \mathcal{H}-\varepsilon \right) ^{-1}\!.
\]%
Then, we find that $G_{11}=\left( 1-G_{11}^{\left( 0\right)
}H_{12}G_{22}^{\left( 0\right) }H_{21}\right) ^{-1}G_{11}^{\left( 0\right) }$%
, so that $G_{11}^{-1}+\varepsilon =H_{11}-H_{12}G_{22}^{\left(
0\right) }H_{21}$. Since $\left\vert \varepsilon \right\vert \ll
\gamma _{1}$, we expand $G_{22}^{\left( 0\right) }=\left(
H_{22}-\varepsilon \right) ^{-1}$ in $\gamma _{1}^{-1}$, keeping
only terms up to quadratic in $\mathbf{p}$ (and therefore in ${\pi
}^{\dag },\pi $), and arrive at the expression in
Eq.~(\ref{heff1}).

\begin{figure}[t]
\centerline{\epsfxsize=1.0\hsize\epsffile{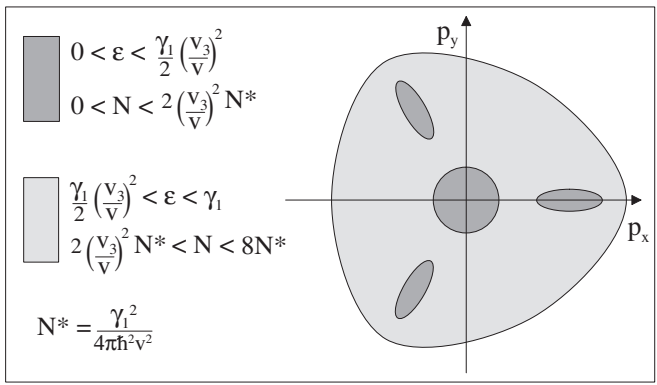}}
\caption{Schematic of the Fermi line in the valley $K$,
$\protect\xi =1$, for high (light shading) and low density (dark
shading). Note that the asymmetry of the Fermi line at valley
$\tilde{K}$, $\protect\xi =-1$, is inverted.} \label{fig:2}
\end{figure}

For low quasiparticle energies, $\left\vert \varepsilon \right\vert \ll
\gamma _{1}$, the spectrum determined by ${\hat{H}}_{2}$ in Eq.~(\ref{heff1}%
) agrees with $\epsilon _{1}\left( p\right) $ found using the $4\times 4$
Hamiltonian $\mathcal{H}$. Similarly to bulk graphite \cite%
{dressel02,dressel74}, the effect of ${\hat{h}}_{w}$ consists of trigonal
warping, which deforms the isoenergetic lines along the directions $\varphi
=\varphi _{0}$, as shown in Fig.~\ref{fig:2}. For the valley $K$, $\varphi
_{0}=0$, ${\textstyle\frac{2}{3}}\pi $ and ${\textstyle\frac{4}{3}}\pi $,
whereas for $\tilde{K}$, $\varphi _{0}=\pi $, ${\textstyle\frac{1}{3}}\pi $
and ${\textstyle\frac{5}{3}}\pi $. At the lowest energies $\left\vert
\varepsilon \right\vert <{\textstyle\frac{1}{2}}\gamma _{1}\left(
v_{3}/v\right) ^{2}$, trigonal warping breaks the isoenergetic line into
four pockets, which can be referred to as one \textquotedblleft
central\textquotedblright\ and three \textquotedblleft
leg\textquotedblright\ parts \cite{dressel74}. The central part and leg
parts have minimum $|\varepsilon |={\textstyle\frac{1}{2}}u$ at $p=0$ and at
$|p|=\gamma _{1}v_{3}/v^{2}$, angle $\varphi _{0}$, respectively. For $%
v_{3}\sim 0.1v$, we find (using the data in Ref. \cite{dressel02}) that the
separation of a 2D Fermi line into four pockets would take place for very
small carrier densities $N<N_{c}=2(v_{3}/v)^{2}N^{\ast }\sim 1\times
10^{11}cm^{-2}$. For $N<N_{c}$, the central part of the Fermi surface is
approximately circular with area $\mathcal{A}_{\mathrm{c}}\approx \pi
\varepsilon ^{2}/(\hbar v_{3})^{2}$, and each leg part is elliptical with
area $\mathcal{A}_{\mathrm{l}}\approx {\textstyle\frac{1}{3}}\mathcal{A}_{%
\mathrm{c}}$. This determines the following sequencing of the first few LL's
in a low magnetic field, $B\ll B_{c}\approx hN_{c}/4e\sim 1T$. Every third
Landau level from the central part has the same energy as levels from each
of the leg pockets, resulting in groups of four degenerate states. These
groups of four would be separated by two non-degenerate LLs arising from the
central pocket.

\begin{figure}[t]
\centerline{\epsfxsize=1.0\hsize\epsffile{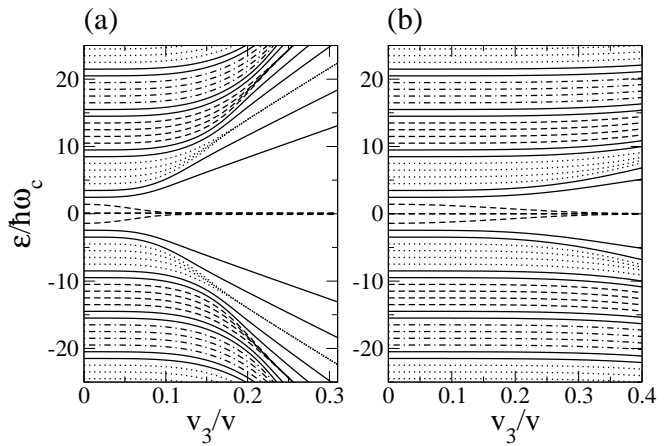}}
\caption{Numerically calculated Landau levels for various values
of $v_{3}$ for valley $K$, $\protect\xi =1$. (a) for $B=0.1T$
($\hbar \protect\omega_{c}=0.216meV$ assuming
$\protect\gamma_{1}=0.39eV$, $v=8.0\times 10^{5}m/s$), (b) for
$B=1T$ ($\hbar \protect\omega_{c}=2.16meV$). Broken lines show
groups of four consecutive levels that become degenerate at large
$v_{3}$, with each group separated from the next by two solid
lines representing levels that are not degenerate.} \label{fig:3}
\end{figure}

In structures with densities $N>N_{c}$ or for strong magnetic fields $%
B>B_{c} $, the above described LL spectrum evolves into an almost
equidistant staircase of levels. We derive such a spectrum numerically from
Eq.~(\ref{heff1}) using the Landau gauge $\mathbf{A}=\left( 0,Bx\right) $,
in which operators ${\pi }^{\dag }$ and $\pi $ coincide with raising and
lowering operators \cite{p+r87} in the basis of Landau functions $%
e^{iky}\phi _{n}(x)$, such that $\pi ^{\dag }\phi _{n}=i(\hbar /\lambda _{B})%
\sqrt{2(n+1)}\phi _{n+1}$, $\pi \,\phi _{n}=-i(\hbar /\lambda _{B})\sqrt{2n}%
\phi _{n-1}$, and $\pi \,\phi _{0}=0$, where $\lambda _{B}=\sqrt{\hbar /(eB)}
$. In this we followed an approach applied earlier to bulk graphite \cite%
{dressel74} and used the bulk parameters for intralayer $v$ and interlayer $%
\gamma _{1}$, varying the value of the least known parameter $v_{3}$. The
spectrum for the valley $K$ ($\xi =1$) is shown in Fig.~\ref{fig:3} as a
function of the ratio $v_{3}/v$ for two different fields. Fig.~\ref{fig:3}%
(a) shows the evolution of the twenty lowest levels for $B=0.1T$ as a
function of $v_{3}$, illustrating the above-mentioned crossover from an
equidistant ladder at $v_{3}=0$ to groups of pocket-related levels.

The LL spectrum obtained for $B=1T$, Fig.~\ref{fig:3}(b) remains independent
of $v_{3}$ over a broad range of its values. Hence, even in the absence of a
definite value of $v_{3}$, we are confident that the LL spectrum in bilayers
studied over the field range where $\hbar \lambda _{B}^{-1}>v_{3}m$ can be
adequately described by neglecting $v_{3}$, thus using an approximate
Hamiltonian given by the first term in ${\hat{H}}_{2}$, Eq.~(\ref{heff1}).
The resulting spectrum contains almost equidistant energy levels which are
weakly split in valleys $K$ ($\xi =+1$) and $\tilde{K}$ ($\xi =-1$),
\begin{eqnarray}
\varepsilon _{n}^{\pm } &=&\pm \hbar \omega _{c}\sqrt{n(n-1)}-{\textstyle%
\frac{1}{2}}\xi \delta ,\;\;\text{for }n\geq 2,  \label{LLhigh} \\
\Phi _{n\xi } &\equiv &C_{n\xi }\left( \phi _{n},D_{n\xi }\phi _{n-2}\right)
,\;\;\;\delta =u\hbar \omega _{c}/\gamma _{1}.  \nonumber
\end{eqnarray}%
Here, $\omega _{c}=eB/m$, $\varepsilon _{n}^{+}$\ and $\varepsilon _{n}^{-}$%
\ are assigned to electron and hole states, respectively, and $D_{n\xi
}=[\varepsilon -\xi u/2+\xi n\delta ]/(\hbar \omega _{c}\sqrt{n(n-1)})$, $%
C_{n\xi }=1/\sqrt{1+|D_{n\xi }|^{2}}$. In the limit of valley ($u=0$) and
spin degeneracies \cite{footnoteZeeman}, we shall refer to these states as
4-fold degenerate LLs.

The LL spectrum in each valley also contains two levels identified using the
fact that ${\pi ^{2}}\phi _{1}={\pi ^{2}}\phi _{0}=0$,
\begin{equation}
\left\{
\begin{array}{ll}
\varepsilon _{0}={\textstyle\frac{1}{2}}\xi u\,; & \quad \Phi _{0\xi }\equiv
(\phi _{0},0); \\
\varepsilon _{1}={\textstyle\frac{1}{2}}\xi u-\xi \delta \,; & \quad \Phi
_{1\xi }\equiv (\phi _{1},0).%
\end{array}%
\right.  \label{LLlow}
\end{equation}%
According to different definitions of two-component $\Phi $ in two valleys, $%
n=0,1$ LL states in the valley $K$ are formed by orbitals predominantly on
the $A$ sites from the bottom layer, whereas the corresponding states in the
valley $\tilde{K}$ are located on $\tilde{B}$ sites from the top layer,
which is reflected by the splitting $u$ between the lowest LL in the two
valleys. In a symmetric bilayer ($u=0$) levels $\varepsilon _{0}$\ and $%
\varepsilon _{1}$\ are degenerate and have the same energy in valleys $K$
and $\tilde{K}$, thus forming an 8-fold degenerate LL at $\varepsilon =0$
(here, spin is taken into account). Also, note that the spectrum of
high-energy LLs, Eq. (\ref{LLhigh}) is applicable in such fields that $\hbar
\lambda _{B}^{-1}<\gamma _{1}/2v$. \ For higher fields the full two-band
Hamiltonian $\mathcal{H}$ has to be used to determine the exact LL spectrum,
nevertheless, the 8-fold degeneracy of the zero-energy LL remains unchanged.

\begin{figure}[t]
\centerline{\epsfxsize=1.0\hsize\epsffile{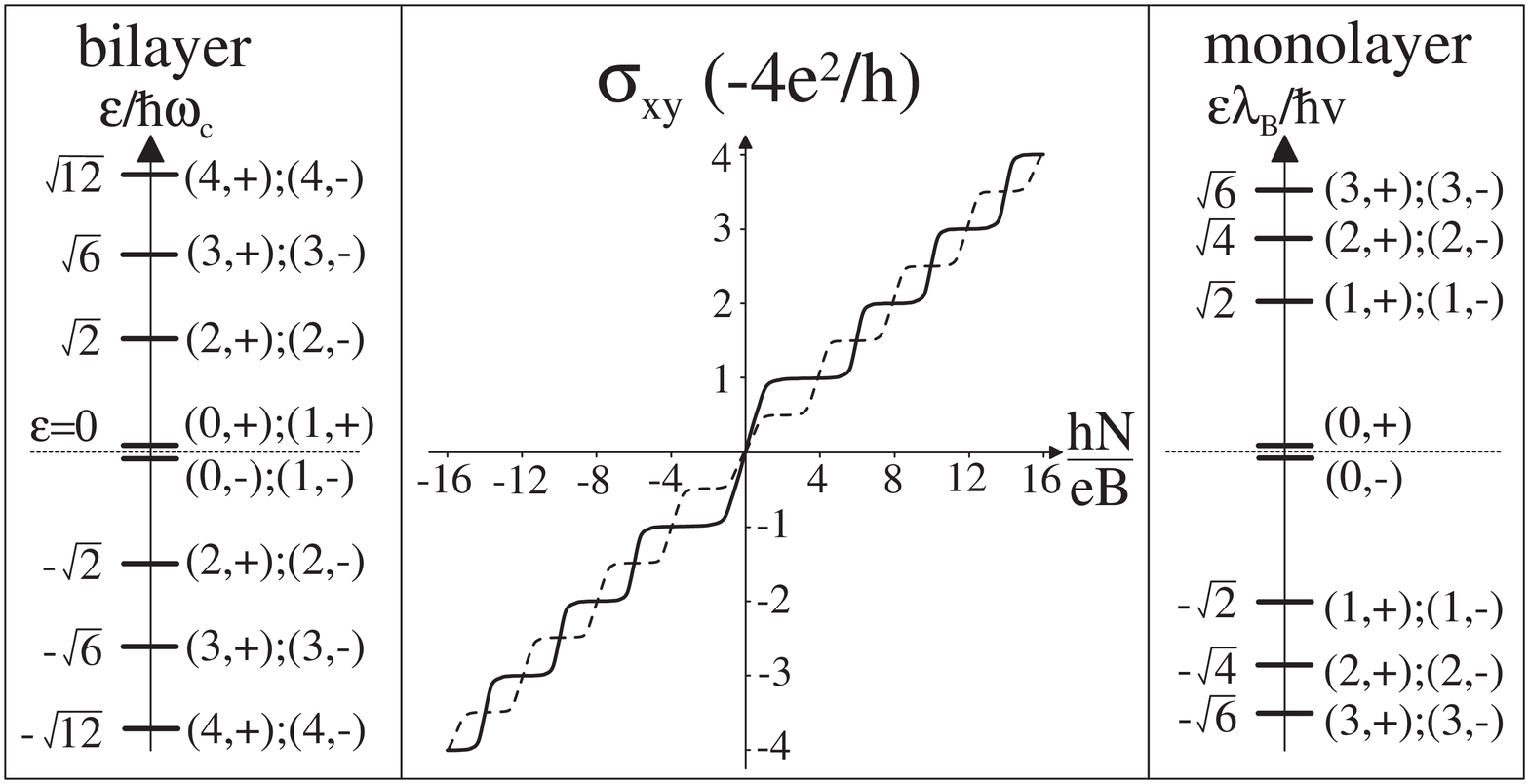}}
\caption{Landau levels for a bilayer (left) and monolayer (right). Brackets $%
(n,\protect\xi )$ indicate LL number $n$ and valley index $\protect\xi =\pm
1 $. In the center the predicted Hall conductivity $\protect\sigma _{xy}$
(center) as a function of carrier density for bilayer (solid line) is
compared to that of a monolayer (dashed line).}
\label{fig:4}
\end{figure}

The group of 8 states at $\left\vert \varepsilon \right\vert =0$ (4 for
electrons and 4 for holes, Eq.(\ref{LLlow})) embedded into the ladder of
4-fold degenerate LL's with $n\geq 2$, Eq. (\ref{LLhigh}) is specific to the
magneto-spectrum of $J=2$ chiral quasiparticles. It would be reflected by
the Hall conductivity dependence on carrier density, $\sigma _{xy}(N)$ shown
in Fig.~\ref{fig:4}. A solid line sketches the form of the QHE $\sigma
_{xy}^{(2)}(N)$ in a bilayer which exhibits plateaus at integer values of $%
4e^{2}/h$ and has a \textquotedblleft double\textquotedblright\ $8e^{2}/h$
step between the hole and electron gases across $N=0$ that would be
accompanied by a maximum in $\sigma _{xx}$. Figure~\ref{fig:4} is sketched
assuming that temperature and the LL broadening hinder small valley and spin
splittings as well as the splitting between $n=0,1$ electron/hole LL's in
Eqs.~(\ref{LLlow}), so that the percolating states \cite{p+r87} from these
levels would not be resolved. To compare, a monolayer has a spectrum
containing 4-fold (spin and valley) degenerate LLs \cite{mcc56}, $%
\varepsilon _{0}=0$ and $\varepsilon _{n\geq 1}^{\pm }=\pm \sqrt{2n}\hbar
v/\lambda _{B}$ shown on the r.h.s of Fig.~\ref{fig:4}, which corresponds to
Hall conductivity $\sigma _{xy}^{(1)}(N)$ exhibiting plateaus at $%
(4n+2)e^{2}/h$ (dotted line \cite{twolayers}), as discussed in earlier
publications \cite{haldane88}.

The absence of a $\sigma _{xy}=0$ plateau in the QHE accompanied
by the maximum in $\sigma _{xx}$ in the vicinity of zero density
is the result of the existence of the zero-energy LL, which is the
fingerprint of a chiral nature of two-dimensional quasiparticles.
This contrasts with a gradual freeze-out of both Hall and
dissipative conductivities in semicondutor structures upon their
depletion. Having compared various types of density dependent Hall
conductivity, we suggest that two kinds of chiral (Berry phase
$J\pi$) quasiparticles specific to monolayer ($J=1$) and bilayer
($J=2$) systems can be distinguished on the basis of QHE
measurements. It is interesting to note that the recent Hall
effect study of ultra-thin films by Novoselov \textit{et al.}
\cite{novo05} featured both types of $\sigma _{xy}(N)$ dependence
shown in Fig.~\ref{fig:4}.

It is also worth mentioning that the 8-fold degeneracy of the
group of $\epsilon =0$ LLs in a bilayer, Eqs.~(\ref{LLlow}), is
quite unusual in 2D systems. It suggests that e-e interaction in a
bilayer may give rise to a variety of strongly correlated QHE
states. For structures studied in Ref.~\onlinecite{novo05}, with
electron/hole densities $N\sim 10^{12}cm^{-2}$, such a regime may
be realized in fields $B\sim 10T$.

The authors thank A.Geim, P.Kim, K.Novoselov, and I.Aleiner for useful
discussions, and EPSRC for support.


\begin{thebibliography}{99}
\bibitem{kpoints} Corners of the hexagonal Brilloin zone are $\mathbf{K}%
_{\xi }=\xi ({\textstyle\frac{4}{3}}\pi a^{-1},0)$, where $\xi =\pm 1$ and $%
a $ is the lattice constant.
\bibitem{footnoteSpaceRev} For spatial inversion we use $(\Pi _{1}\otimes
\sigma _{0}){\hat{H}}\left( {\mathbf{p},B,{u}}\right) (\Pi _{1}\otimes
\sigma _{0})={\hat{H}}\left( -{\mathbf{p},B,-{u}}\right) $ where $\Pi _{1}$
swaps $\xi =\pm 1$ valleys.
\bibitem{footnoteZeeman} Although in graphite the electron g-factor is not
small $g=2$, a very light effective mass $m=\gamma_{1}
/2v^{2}\approx 0.054m_{e}$ in the bilayer determines a small ratio
between the Zeeman energy and LL splitting, $\varepsilon_{Z}
/\hbar \omega _{c}\sim 0.05$.
\bibitem{twolayers} Note that two parallel equivalent monolayers would
display Hall conductivity 2$\sigma _{xy}^{(1)}(N)$, thus missing
every second plateau, as compared to $\sigma _{xy}^{(2)}(N)$.
\end{thebibliography}
\end{document}